 \documentclass[letterpaper,twocolumn,showpacs
,prb
,citeautoscript
,superscriptaddress%
]{revtex4}

\usepackage{graphicx}
\usepackage{amssymb}
\usepackage{amsmath,amsfonts,latexsym}
\usepackage{array,tabularx,color}
\usepackage{dcolumn}               

\DeclareMathOperator{\Veff}{V^{\text{eff}}_{\vect{k},\vect{k}',\vect{q}}}
\DeclareMathOperator{\ex}{x}

\newcommand{\vect}[1]{{\mathbf #1}}
\newcommand{\Frac}[2]{\displaystyle\frac{#1}{#2}}

\newcommand{\eb}{E_{\text{XX}}}

\begin{document}
\title{Phase diagram for condensation of microcavity polaritons: from
  theory to practice}
\author{F.~M.~Marchetti}
\affiliation{Rudolf Peierls Centre for Theoretical Physics, 1 Keble
Road, Oxford OX1 3NP, UK}
\author{M.~H.~Szyma{\'n}ska}
\affiliation{Department of Physics, University of Warwick, Coventry CV4 7AL, UK}
%
\author{J.~M.~J.~Keeling}
\affiliation{Cavendish Laboratory, University of Cambridge,
             Madingley Road, Cambridge CB3 0HE, UK}
\author{J.~Kasprzak}
\altaffiliation{Present address: School of Physics and Astronomy,
Cardiff University, Queens Buildings, 5 The Parade, Cardiff, CF24 3AA, UK}
\affiliation{CEA-CNRS-UJF joint group `Nanophysique et
  Semiconducteurs', Institut Ne\'el, CNRS, 25 rue des Martyrs, 38042
  Grenoble, France}
%
\author{R. Andr\'e}
\affiliation{CEA-CNRS-UJF joint group `Nanophysique et
  Semiconducteurs', Institut Ne\'el, CNRS, 25 rue des Martyrs, 38042
  Grenoble, France}
%
\author{P.~B.~Littlewood} 
\affiliation{Cavendish Laboratory, University of Cambridge, Madingley
             Road, Cambridge CB3 0HE, UK}

\author{Le Si Dang}
\affiliation{CEA-CNRS-UJF joint group `Nanophysique et
  Semiconducteurs', Institut Ne\'el, CNRS, 25 rue des Martyrs, 38042
  Grenoble, France}

\date{March 23, 2007}       

\begin{abstract}
  The first realization of a polariton condensate was recently
  achieved in a CdTe microcavity [Kasprzak \emph{et al.}, Nature
  \textbf{443}, 409 (2006)].  We compare the experimental phase
  boundaries, for various detunings and cryostat temperatures, with
  those found theoretically from a model which accounts for features
  of microcavity polaritons such as reduced dimensionality, internal
  composite structure, disorder in the quantum wells,
  polariton-polariton interactions, and finite lifetime.
\end{abstract}

\pacs{71.35.Lk, 03.75.Gg, 71.36.+c}            

\maketitle

\section{Introduction}
\label{sec:introduction}

Microcavity polaritons~\cite{hopfield58,weisbuch92} are the
quasi-particles which result from strong light-matter coupling in
semiconductor microcavities. The light polariton mass means a weakly
interacting dilute Bose gas picture would imply a high transition
temperature~\cite{kavokin03:pla,malpuech03,keeling06} and, even with
more sophisticated
methods~\cite{keeling04:polariton,marchetti06:prl,sarchi08}, the
expected transition temperature is $10^8$ times higher than for atomic
gases. While Bose-Einstein condensates (BEC) of atomic gases are now
routinely studied, condensation of microcavity polaritons has been
long
sought~\cite{dang98:prl,deng02,deng03,richard04:jpcm,richard05,richardprb05,deng06:eqbm}
and only recently has a thermal equilibrium condensate accompanied by
spontaneous coherence been realized in a CdTe
microcavity~\cite{kasprzak06:nature}. In the same structure the
polarisation has been investigated\cite{kasprzak07}, second order
temporal coherence has been studied\cite{kasprzak08}, and pinned
vortices have been observed in the condensed
phase~\cite{lagoudakis08}. Spontaneous coherence of microcavity
polaritons has been also recently investigated in
GaAs~\cite{deng07,snoke07science}, while there have been early reports
on room temperature polariton lasing in a GaN
cavity~\cite{christopoulos07}.

Microcavity polaritons are interacting particles in a disordered
potential with internal structure and finite lifetime, which form a
finite two-dimensional gas, so their condensation differs from that of
ideal non-interacting three-dimensional (3D) bosons for which BEC was
originally discussed.  It is therefore crucial to establish these
differences and the nature of polariton condensation.
To this end, the aim of this paper is to compare theoretical and
experimental results for the phase diagram, and to discuss what they
reveal about the nature of the polariton condensate.
We discuss both complete thermal equilibrium phase diagrams, and also
consider effects that may result from pumping and decay (even when the
polariton distribution is well thermalised).
As a further way of illustrating the differences between the ideal 3D
Bose gas and microcavity polaritons, the condensate fraction as a
function of pumping strength is presented; this is compared
qualitatively to a theoretical model including effects of pumping and
decay.

We experimentally evaluate the critical density for condensation at
low cryostat temperatures in the range $T_{\text{cryo}}=(5,15)$K and
for exciton-photon detunings in the range $\delta=(5.3,12.5)$ meV.
These densities are compared with the predictions of a realistic
microscopic model that accounts for the composite structure of
microcavity polaritons, interactions enhanced by quantum well disorder
and finite polariton lifetime.
This comparison reveals two crucial aspects of the current realisation
of polariton condensation in CdTe: that it lies in the crossover
between BEC of weakly interacting bosons and the mean-field phase
transition driven by interactions; and that the excitonic disorder is
important in the evaluation of the polariton-polariton interactions.

The organisation of the remainder of this paper is as follows:
Section~\ref{sec:exper-determ-crit} describes the experimental
estimate of critical density, found by integrating the occupation.
Section~\ref{sec:theor-phase-diagr} then introduces the theoretical
models we consider, and in Sec.~\ref{sec:comp-exper-theor} we discuss
the various theoretical estimates of phase boundaries they produce.
The comparison between the various estimates of experimental critical
density and the theoretical phase boundary is shown in
Sec.~\ref{sec:disc-poss-errors}.
The effects of pumping and decay are discussed in
Sec.~\ref{sec:effects-pump-decay},  and their influence on the
condensate fraction -- in Sec.~\ref{sec:condensate-fraction}.

\section{Experimental determination of critical density}
\label{sec:exper-determ-crit}
\begin{figure}[htpb]
\begin{center}
\includegraphics[width=0.9\linewidth,angle=0]{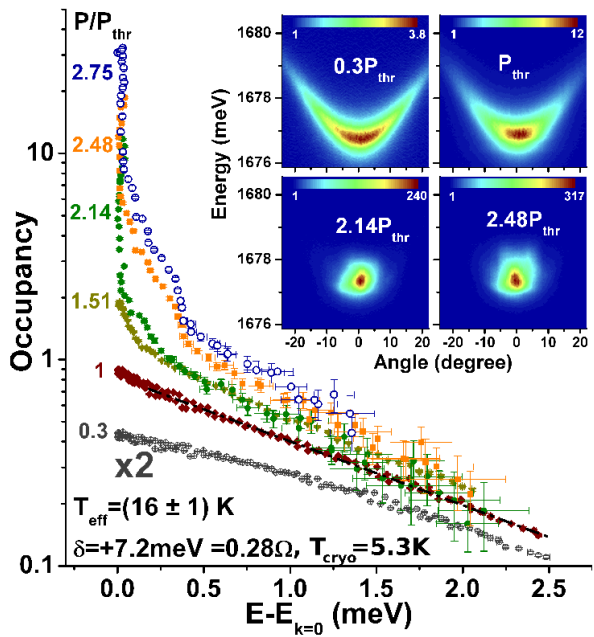}
\end{center}
\caption{(Color online) LP occupation versus energy at various
  excitation powers extracted from the energy-resolved far-field
  emission (inset).}
\label{fig:occup}
\end{figure}

A CdTe/CdMgTe microcavity with 16 quantum wells and Rabi splitting
$\Omega_R=26$meV, is non-resonantly pumped at the first high energy
lobe of the Bragg structure, at $1.768$eV (about $\sim 100$meV above
the lower polariton ground state energy) by a continuous-wave
Ti:sapphire laser combined with an acousto-optic modulator ($1\mu$s
pulse duration with a duty cycle of $1\%$) to reduce the sample
heating.  The excitation spot is uniform and of $\sim 35\mu$m
diameter.  Following Ref.~\onlinecite{kasprzak06:nature}, we detune
the photon energy $\omega_{\vect{k}}$ above the exciton $E_{\ex}$
($\delta=\omega_{0} - E_{\ex}$) and, increasing the pump laser power,
we measure the energy- and angle- (or momentum-) resolved emission
intensities via far-field spectroscopy (inset of
Fig.~\ref{fig:occup}). In-plane momentum $\vect{k}$ and emission angle
$\theta_{\vect{k}}$ are related by $\hbar c |\vect{k}| =
E^{\text{LP}}_{\vect{k}} \sin \theta_{\vect{k}}$, where
\begin{displaymath}
 E^{\text{LP,UP}}_{\vect{k}} = \frac{1}{2}[\omega_{\vect{k}} + E_{\ex} \mp
\sqrt{(\omega_{\vect{k}} - E_{\ex})^2 + \Omega_R^2}]
\end{displaymath}
are the lower (LP) and upper polariton (UP) energies. From the
energy-integrated photoluminescence (PL) at a given momentum
$\mathcal{P}_{\vect{k}}$, the occupation $\mathcal{O}_{\vect{k}}$ of
that state is given by~\cite{ciuti03}
\begin{displaymath}
  \mathcal{O}_{\vect{k}} = 
  \frac{\mathcal{P}_{\vect{k}}}{\cos^4 \theta_{\vect{k}} c_{\vect{k}}^2}
  , \quad
  c_{\vect{k}}^2 = \frac{1}{2}
  \left[1 -
    \frac{\omega_{\vect{k}} - E_{\ex}}{\sqrt{(\omega_{\vect{k}} - E_{\ex})^2 +
        \Omega_R^2}}
    \right]\; .  
\end{displaymath}
By finding the energy at which, for a given momentum, the PL is
maximum, momentum dependence can be converted to energy dependence; an
example is shown in Fig.~\ref{fig:occup}. Above a threshold pump
power, the ground state occupation grows exponentially, while the
effective LP temperature, $T_{\text{eff}}$, (extracted from the tail
of the occupation) does not change much. This macroscopic occupation
of the ground state has been shown to be associated with macroscopic
coherence across the spot size~\cite{kasprzak06:nature}, demonstrating
polariton condensation.  At threshold $T_{\text{eff}}=16$K; this is
above $T_{\text{cryo}}$ because the polariton lifetime (of $1$--$2$ps)
is shorter than typical relaxation times (although longer than their
thermalization time~\cite{kasprzak06:nature,deng06:eqbm}).
The ground state occupation at threshold is, within a factor of two,
measured to be one; in Fig.~\ref{fig:occup} this is rescaled to one.

The LP density at the threshold pump power can then be estimated by
integrating the occupation to give the total density.
The estimate found this way is represented on Fig.~\ref{fig:phas1}
by blue stars.
Since the system is thermalized, by rescaling the occupation at
$\vect{k}=0$ to one, this density estimate is determined by the
effective temperature only; any variation at fixed temperature is due
to changes to the LP density of states.

\section{Theoretical  phase diagrams}
\label{sec:theor-phase-diagr}

This section discusses the theoretical modelling of the phase diagram,
and further methods of estimating the experimentally observed critical
density.
The discussion here concerns the thermal equilibrium case; possible
effects of pumping and decay are discussed later in
Sec.~\ref{sec:effects-pump-decay}.
The discussion will compare the use of two models; the first --- model
\eqref{eq:hamil} --- is a model of disorder localised saturable
excitons coupled to propagating
photons~\cite{marchetti06:prl,marchetti07}.
At low enough densities and temperatures, the critical density given
by model~\eqref{eq:hamil} will be dominated by long wavelength bosonic
fluctuations~\cite{keeling04:polariton,keeling05}, and so the results
of this model can be reproduced by modelling lower polaritons as a
weakly interacting dilute Bose
gas~\cite{ciuti03,keeling06,keeling_review07}, which we shall call
model \eqref{eq:lowmo}.
Excitonic disorder can significantly modify the effective 
polariton interaction compared to the clean case.
To take account of excitonic disorder in model~\eqref{eq:lowmo}, we
choose the interaction strength for this model so as to match the 
low density limit of model~\eqref{eq:hamil}, in which excitonic 
disorder is explicitly considered.
Such matching is achieved by calculating the form of blueshift vs
density in both models, and choosing the effective interaction
strength for the bosonic model so they agree.
This calculation of blueshift vs density also allows a second estimate
of the experimentally found critical density, by measuring the
blueshift observed at that density.

\subsection{Localised exciton model}
\label{sec:local-excit-model}

The first model, following
Refs.~\onlinecite{marchetti06:prl,marchetti07}, starts from the
single-particle exciton states (numerically) evaluated in a disordered
quantum well, as explained further below.
Although being 2D, all such states are localised, the nature of the
exciton states and their associated oscillator strength change
substantially with the exciton energy~\cite{runge98}.
The exciton states that couple most strongly to the cavity photons are
those just below the band edge.

Although the exciton states are localised, polaritons consist of a
superposition of many excitons~\cite{eastham00:ssc} and so need not be
localised~\cite{whittaker00_RRS}. 
Polaritons in CdTe are also localised by photonic disorder, associated
with the spatial inhomogeneity observed in the PL above threshold.
However, above threshold, coherence can be observed over a large
length scale approaching the excitation spot
size~\cite{kasprzak06:nature,baas07}, and so polaritons could not be
localised on a shorter length scale.
This gives an energy scale up to $0.1$meV$\simeq1$K, much smaller than
the observed blue-shift (due to interaction). 
This difference in energy scales suggests that localisation is a
perturbation to a theory of interacting bosons, rather than vice
versa; we therefore neglect photonic disorder when finding the phase
boundary.

\subsubsection{Many body Hamiltonian}
\label{sec:many-body-hamilt}

In the many body Hamiltonian, we approximate the
interactions between excitons by exclusion: Each single-particle 
exciton state can only be occupied once.
This approximation over-estimates the on-site interaction and neglects
the inter-site interaction --- in reality there is finite Coulomb and
exchange energy to multiply occupy the same single exciton states.
However, as discussed in Ref.~\onlinecite{keeling_review07}, this
approximation is valid at low enough densities.
This procedure yields model \eqref{eq:hamil},
\begin{multline}
  \hat{H} = \sum_{\alpha} \varepsilon_{\alpha} S_{\alpha}^{z} +
    \sum_{\vect{k}} \omega_{\vect{k}} \psi_{\vect{k}}^\dag
    \psi_{\vect{k}}^{} \\ + \Frac{1}{\sqrt{A}} \sum_{\alpha, \vect{k}}
    \left(g_{\alpha , \vect{k}} \psi_{\vect{k}}^{} S_{\alpha}^{+} +
    \text{h.c.}\right) \; .
\label{eq:hamil}
\end{multline}
The spin operators $S^{z}, S^{\pm}$ describe states
$|\!\!\uparrow_{\alpha} \rangle$ and $|\!\!\downarrow_{\alpha}
\rangle$ corresponding respectively to the presence or absence of an
exciton of energy $\varepsilon_\alpha$.
The operators $\psi_{\vect{k}}^{}$ describe confined photon modes of
energy $\omega_{\textbf{k}} = \omega_0 + \hbar^2 \mathbf{k}^2/2
m_{\text{ph}}$ and $m_{\text{ph}}/m_0=2.58\times 10^{-5}$ ($m_0$ is
the bare electron mass).
The quantisation area $A$ that appears in Eq.~\eqref{eq:hamil} plays
no role in any final answer, and is taken to infinity in all our
calculations.
The Hamiltonian written in Eq.~\eqref{eq:hamil} does not yet take
account of spin of the exciton states or polarisation states of the
light.
To a first approximation, the interaction between different spins is
weak; in such a limit each polarisation behaves separately, and one
has independent copies of this model for each spin, thus introducing a
degeneracy factor $g_s=2$ in the total density.
Beyond this limit, opposite spin polarisations are weakly attractive,
producing bound excitonic states; the effect of such an interaction is
discussed in Appendix~\ref{sec:inter-betw-spins}.

\subsubsection{Energies and coupling strength of excitons}
\label{sec:energ-coupl-strength}

As in Refs.~\onlinecite{marchetti06:prl,marchetti07}, the exciton
energies $\varepsilon_{\alpha}$ and coupling strengths
$g_{\alpha,\vect{k}}$ (which lead to the inhomogeneous exciton
linewidth) are taken from numerical diagonalisation of exciton
wavefunctions in a disorder potential~\cite{runge98}.
These numerical calculations are described in detail in
Ref.~\onlinecite{marchetti07}, so we only briefly summarise the method
here.
We numerically solve the Schrodinger equation in a disorder potential:
\begin{displaymath}
  \left[
    - \frac{\nabla_\vect{R}^2}{2 m_{\ex}} + V(\vect{R}) + E_{\ex} 
  \right]
  \Phi_{\alpha}(\vect{R})
  =
  \varepsilon_\alpha \Phi_{\alpha}(\vect{R})\; .
\end{displaymath}
We use the CdTe exciton mass~\cite{yakovlev97} $m_{\ex}/m_0=0.6$,
and a Gaussian correlated exciton disorder.
For disorder weaker than the exciton binding energy, the internal
structure of the exciton smooths disorder on lengthscales shorter than
the exciton Bohr radius~\cite{runge98} $a_{\ex}$, so we can 
consider a Gaussian noise correlated on a length scale $\ell_c$:
\begin{displaymath}
  \langle V(\vect{R}) \rangle=0, \quad
  \left< V(\vect{R}) V(\vect{R}^{\prime}) \right>
  =
  \frac{\sigma^2 \ell_c^2}{A} 
  \sum_\vect{q}^{1/\ell_c}
  e^{i\vect{q}\cdot(\vect{R}-\vect{R}^{\prime})}\; .
\end{displaymath}
We take $\ell_c = 167 \text{\AA} \gtrsim a_{\ex}$ and so the only free
parameter in this calculation is the strength $\sigma$ of the
disorder; this was chosen to give an exciton linewidth that matches
the observed CdTe exciton linewidth ($\sim 1$meV), leading to
$\sigma=0.79$meV.

The exciton-photon coupling strength involves the overlap between the
disorder localised exciton state, and the photon wavefunction.
Since the relevant photon momenta are small compared to the typical
scales of the exciton wavefunction (i.e. small compared to
$1/a_{\ex}$), we take this coupling strength at zero photon
momentum, as $g_{\alpha,\vect{k}} \simeq g_{\alpha,0} \propto \left<
  \Phi_{\alpha} | \vect{k}=0 \right>$, where $\Phi_{\alpha}$ is the
disorder localised exciton state.
To fix the overall scale of the couplings we use Rabi splitting
$\Omega_R = 2 \sqrt{\sum_{\alpha} |g_{\alpha, 0}|^2/A}$ with the
observed Rabi splitting.
Thus, the complete set parameters defining model~\eqref{eq:hamil} are
the exciton mass, the Rabi splitting, the exciton linewidth, the
exciton Bohr radius and the photon mass.
These parameters all have reasonably well established values, and
are not used as fitting parameters.

\subsection{Weakly interacting dilute Bose gas}
\label{sec:weakly-inter-dilute}

At low densities and temperatures, the thermodynamics of model
\eqref{eq:hamil} is dominated by long wavelength bosonic
fluctuations\cite{keeling04:polariton,keeling05}, and so it gives the
same predictions as an effective LP
model~\cite{ciuti03,keeling06,keeling_review07} [model
\eqref{eq:lowmo}]:
\begin{equation}
  \hat{H}\simeq \sum_{\vect{k}} E_{\vect{k}}^{\text{LP}}
  L_{\vect{k}}^\dag L_{\vect{k}}^{} +
  \sum_{\vect{k},\vect{k}',\vect{q}}
  \Veff
  L_{\vect{k}+\vect{q}}^\dag L_{\vect{k}'-\vect{q}}^\dag
  L_{\vect{k}'}^{} L_{\vect{k}}^{} \; .
\label{eq:lowmo}
\end{equation}
As before, in the absence of biexcitonic effects, the spin degree
of freedom only contributes a factor $g_s=2$ to the density.
Such a model for the condensation of weakly interacting bosons is
often studied in the context of excitons, and of atomic
gases~\cite{zzz_bec}.
For polaritons, model~\eqref{eq:lowmo} has been
considered~\cite{kavokin03:pla,malpuech03} with the interaction
strength $\Veff$ taken from a
combination of Coulomb interaction and the nonlinearity of
exciton-photon coupling~\cite{ciuti03}.
However, excitonic disorder can significantly enhance
polariton-polariton interactions~\cite{marchetti06:prl,marchetti07}
and make them 
larger than those in clean QWs.
In choosing the value of $\Veff$ we wish to include these 
effects of disorder, which are described microscopically in
model~\eqref{eq:hamil}.
As mentioned above, the form of model~\eqref{eq:lowmo} describes the
relevant excitations of model~\eqref{eq:hamil} at low densities and
temperatures, and so the physical predictions of these two models
should match in this limit.
Hence, the effect of disorder on the interaction strength can be
included by choosing the value of $\Veff$ so as to make the results 
of model \eqref{eq:hamil} and model \eqref{eq:lowmo} match at low
densities and temperatures.

\subsubsection{Fixing effective interaction strength}
\label{sec:fixing-eff-interaction}

\begin{figure}[htpb]
\begin{center}
  \includegraphics[width=0.9\linewidth,angle=0]{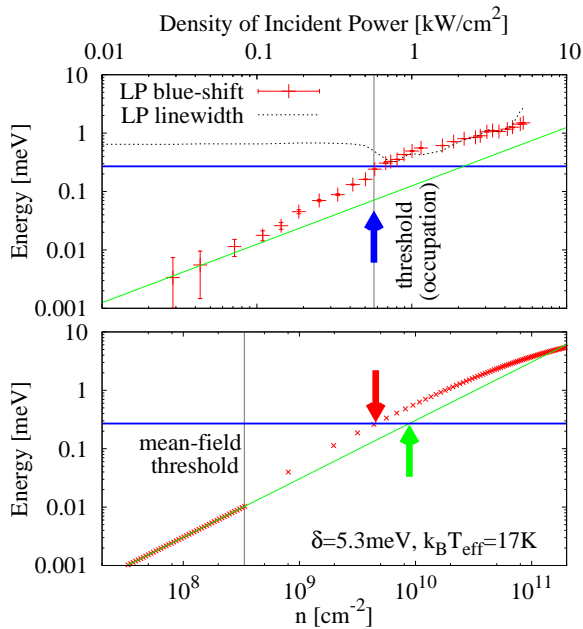}
\end{center}
\caption{(Color online) Comparison of (panel a) experimental LP
  blue-shift (red data with error bars) vs.  incident pump power;
  (bottom panel) theoretical LP blue-shift (red $\times$ symbols)
  vs. mean-field density. The experimental LP linewidth (black dashed)
  and threshold (found from the occupation data) are also
  shown. Measured blue-shifts at threshold can be translated to 
  densities as explained in the text.}
\label{fig:blue1}
\end{figure}

Following the relation between the two models as described above, the
value of $V^{\text{eff}}$ is chosen so that the blueshift vs density
in the normal state calculated in both models will match.
In the normal state of model~\eqref{eq:lowmo}, the LP blue-shift
varies linearly with the density $\delta E^{\text{LP}} \sim
V^{\text{eff}} n$.
In model~\eqref{eq:hamil} the blue-shift depends on both temperature
and disorder, but, in the low density normal state, it also varies
approximately linearly (lower panel of Fig.~\ref{fig:blue1}).
We choose $V^{\text{eff}}$ so these gradients are the same.
In both models, once condensed, the LP branch locks to the chemical
potential, and so the blue-shift traces the dependence of chemical
potential on density, increasing more rapidly then linearly at first,
before eventually saturating (illustrated by the red $\times$ symbols
in the lower panel of Fig.~\ref{fig:blue1}).

The value found by this procedure is
$V^{\text{eff}}=1.52\mu\text{eV}\mu\text{m}^2$, implying a blue-shift
of $0.304$meV at a density of $10^{10}\text{cm}^{-2}$).
This effective interaction strength $V^{\text{eff}}$ is 
roughly 10 times that found from Coulomb and saturation effects in a
clean system~\cite{ciuti03}.
The relative size of inter-site Coulomb and saturation effects can
also be seen in the UP energy shift with pump power: The Coulomb term
shifts both LP and UP in the same direction, saturation effects lead
to opposite shifts.
Experimental data in the on-line supplementary information of
Ref.~\onlinecite{kasprzak06:nature} show an UP red-shift of magnitude
comparable to the LP blue-shift, strongly suggesting that saturation
effects dominate Coulomb interactions.

\subsubsection{Estimating experimental density from blueshift}
\label{sec:estim-exper-dens}

The calculation of blue shift as a function of density allows one to extract a
second estimate of the LP density at threshold, by converting the
experimentally observed blueshift (Fig.~\ref{fig:blue1}, upper panel) 
to the corresponding density.
However, as seen in Fig.~\ref{fig:blue1}, the mean-field estimate of
blue-shift differs according to whether one considers the condensed or
uncondensed case.
Since fluctuations beyond the mean-field approximation delay the
transition, the actual blue-shift at threshold should lie between the
condensed and uncondensed mean-field estimates.
Thus, for a given observed blue shift, one can extract two estimates
of the density that should bound the actual density.
These two translations to density are shown in Fig.~\ref{fig:phas1},
as red upward triangles for the condensed calculation, and green
downward triangles for the uncondensed case.

\subsection{Theoretical phase boundaries}
\label{sec:comp-exper-theor}

\begin{figure}[htpb]
\begin{center}
\includegraphics[width=0.95\linewidth,angle=0]{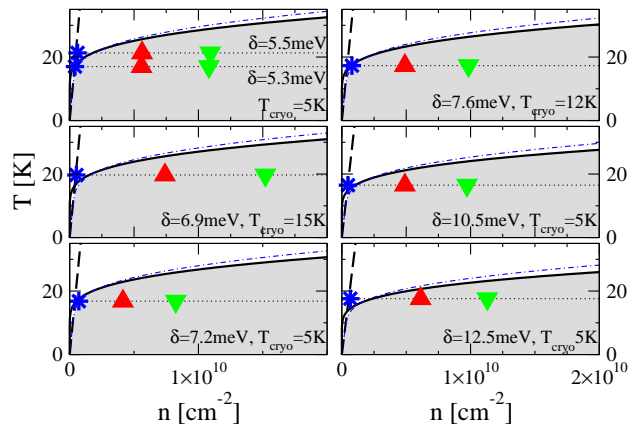}
\end{center}
\caption{(Color online) Theoretical phase diagram of temperature
  versus density (gray region is condensed) for different values of
  the detuning $\delta$ and $T_{\text{cryo}}$, with superimposed 
  the estimates of density from the PL (blue stars) and from the
  blue-shift of LP (red upper and green lower triangles).}
\label{fig:phas1}
\end{figure}

The experimental estimate of the critical polariton density can be now
compared to the theoretical phase diagram. 
From model~\eqref{eq:hamil}, one may derive a mean-field phase
diagram, for condensation of polaritons in the lowest energy state.
The mean-field phase boundary\cite{eastham00:ssc,marchetti07} comes
from finding the mean-field estimate of density ($g_s=2$ takes into
account the degeneracy introduced by the excitonic spin):
\begin{equation}
  \label{eq:two-level-density}
  n = \frac{1}{A} \sum_{\alpha}
  \frac{g_s}{2}\left[
    1  - \tanh\left(\frac{\beta(\varepsilon_\alpha-\mu)}{2}\right)
  \right]\; ,
\end{equation}
at the chemical potential $\mu$ which first satisfies the saddle-point
equation (which plays a role analogous to the self consistency
condition or Gross-Pitaevskii equation for a weakly interacting 
dilute Bose gas):
\begin{equation}
  \label{eq:two-level-gap}
  \omega_0 - \mu = \frac{1}{A}
  \sum_{\alpha} |g_{\alpha,0}|^2 
  \frac{\tanh[\beta (\varepsilon - \mu)/2]}{\varepsilon - \mu}\; .
\end{equation}
This mean-field boundary is shown as a black solid line in
Fig.~\ref{fig:phas1}.

Including fluctuation corrections to this mean-field transition, one
recovers a smooth crossover from the mean-field limit at higher
densities to a fluctuation dominated regime at low
densities~\cite{keeling04:polariton,keeling05}; in this fluctuation
dominated regime, the results of model \eqref{eq:hamil} are equivalent
to those of model \eqref{eq:lowmo}.
This crossover is at a temperature that is controlled by Rabi
splitting, and is of the order of one tenth of the Rabi splitting.
In 2D, for interacting bosons with a quadratic dispersion, the
transition temperature to the superfluid state, the
Berezhinskii-Kosterlitz-Thouless (BKT)~\cite{kosterlitz73,nelson77}
temperature, varies linearly with the density $n$ and is close to the
quantum degeneracy temperature, $T_{\text{deg}}\propto n/m$.
The results of model \eqref{eq:lowmo} are not quite so simple as the
dispersion is non-quadratic~\cite{kavokin03:pla,malpuech03,keeling06}.
However, at ultra-low densities, the phase boundary of
model~\eqref{eq:lowmo} (blue dotted-dashed line) recovers a linear
dependence of critical temperature on density (black dashed line).

The BKT transition discussed above describes an infinite
two-dimensional system.  The experimental system however contains
photonic disorder, which might act as a trapping potential for
polaritons.  In a non-interacting harmonically trapped two-dimensional
Bose gas, a transition exactly like that described by Bose and
Einstein can exist due to the modification of density of states by the
harmonic trap~\cite{bagnato91,ketterle96}.  However, the polariton
system is interacting, and it is not a-priori valid to ignore
interactions; the inclusion of interactions in an harmonically trapped
system~\cite{holzmann07} can replace the BEC transition of the
non-interacting gas with a transition better described by BKT physics.
The exact details of the critical behaviour in a trapped
two-dimensional interacting system are delicate.  As already discussed
in Sec.~\ref{sec:local-excit-model}, for the parameters relevant in
the current system, the effects of polariton trapping induced by the
photonic disorder are weak compared to the polariton-polariton
interaction strength.  However our interest here is in the prediction
of the phase boundary, we note that for the parameters relevant here
simple estimates of the critical temperature for the BEC of
non-interacting trapped system do not differ significantly from those
for the BKT transition.  There is a quite separate question concerning
coherence (and condensate fraction) in a two dimensional trapped
system, which we discuss below in Sec.~\ref{sec:condensate-fraction}.

\section{Comparison of theoretical and experimental boundaries}
\label{sec:disc-poss-errors}

In Fig.~\ref{fig:phas1} the experimental density is near the
crossover between the low- and high-density parts of the theoretical
phase boundary.
The density estimated from the blue shift is slightly greater than
that from integrating the occupation; however without the effects of
excitonic disorder, the density from the blueshift would be ten
times higher.
Let us consider the possible systematic errors involved in
estimating the density.
In integrating occupation, the sources of uncertainty are the
normalisation (discussed earlier), and the accuracy with which the
critical pumping power has been determined.
We may underestimate the experimental critical density if the
nonlinear threshold we use is in fact associated with the diverging
susceptibility on approaching the transition rather than with the
condensation transition itself.
As an upper bound on this source of uncertainty we note that the
density when the linewidth reaches a minimum is between $1.6$ and $2$
times that at threshold.
For the estimates from the blueshift, the uncertainties here are
harder to quantify.
The exclusion interaction for excitons may overestimate blueshift,
implying our estimated density is too low.
Alternatively, our calculation assumes a thermalised polariton
distribution, whereas the influence of the exciton reservoir and hot
carriers remains unclear within the present modelling, these might
increase the blueshift, suggesting that our estimated density is too
high.
Finally, the theoretical phase boundary as yet neglects dephasing;
it is this effect to which we next turn.

\section{Effects of pump and decay}
\label{sec:effects-pump-decay}

As is apparent from Fig.~\ref{fig:occup}, and the discussion in
Ref.\onlinecite{kasprzak06:nature}, the polariton distribution is
thermalised.
However a thermalised population distribution does not necessarily
imply that one can neglect
effects~\cite{szymanska06:prl,szymanska06:long,wouters06b,wouters07,keeling07:gpe,lagoudakis08}
of pumping and decay.
To understand this distinction, let us consider two dimensionless
ratios that characterise non-equilibrium effects.
One might estimate the effect of pumping and decay by considering the
ratio of polariton lifetime to thermalization time (e.g., $\sim 2$ in
the current experiment, $\sim 2000$ in early atomic
gas~\cite{davis95} BECs).
When this ratio is larger than one, the distribution will be
thermalised.
 However, even with a thermal polariton
distribution, pumping and decay cause a particle flux, introducing
dephasing, which increases the critical density at a given
temperature~\cite{szymanska06:prl}.
The importance of this dephasing can be estimated by comparing the
homogeneous photon linewidth, LW$\approx1$meV, to the temperature $k_B
T \approx 2$meV.
As these parameters are comparable, linewidths will have significant
homogeneous contributions, and dephasing can affect the ability of the
system to maintain a coherent state.

\begin{figure}[htpb]
  \begin{center}
    \includegraphics[width=0.9\linewidth,angle=0]{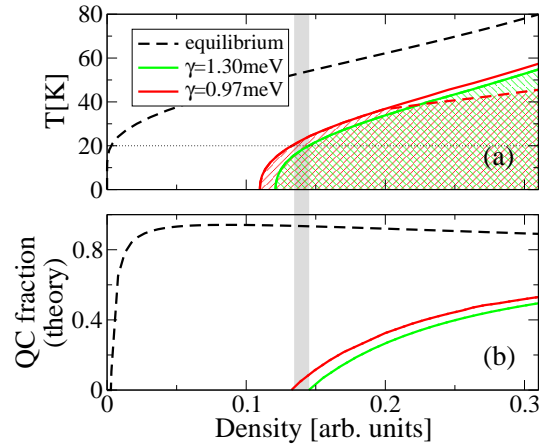}
  \end{center}
  \caption{(Color online) Critical temperature versus density (in
    arbitrary units) (panel a) and quasi-condensate fraction at
    $T=20$K (panel b) including the effects of pumping (rate $\gamma$)
    and decay, with decay rate for a homogeneous cavity line-width of
    $1$meV, for a model Gaussian distribution of the excitonic
    energies.  In panel (a), for $\gamma=0.97$meV and high
    temperatures the nonequilibrium model has an instability of the
    homogeneous condensate\cite{szymanska06:long}; the boundary of
    this instability is marked by the dashed line, and the stable
    condensed region is marked by hatchings.  }
  \label{fig:noneq}
\end{figure}

As an estimate of how such dephasing would modify critical density at
a given temperature, Fig.~\ref{fig:noneq} compares the phase
boundaries found from calculations using model~\eqref{eq:hamil}, but
with photon decay and injection of
excitons~\cite{szymanska06:long,szymanska06:prl} taken into account.
These calculations introduce a flux of particles by allowing photons
to decay and continually injecting new excitons to maintain a steady
state.
The departure from the closed system picture is controlled by
two parameters, the photon decay rate $2\kappa = 1$meV, and a pumping
rate $\gamma$ that describes coupling between the polariton system and
the reservoir of high energy excitons.
The value of $\gamma$ is not clearly known, but is bounded.
For $\gamma \gg \kappa$ one has a polariton linewidth larger than the
bare photon linewidth; for $\gamma \ll \kappa$ there is instability of
the homogeneous condensate\cite{szymanska06:long}; a range of $\gamma$
bounded by these consideration is shown for illustration in
Fig.~\ref{fig:noneq} --- within these limits, the exact value of
$\gamma$ does not significantly shift the curves.
Although these calculations use model~\eqref{eq:hamil}, for
tractability of the nonequilibrium calculations, a Gaussian
distribution of excitonic energies $\varepsilon_{\alpha}$ and a
constant coupling $g_{\alpha,\vect{k}}=\Omega_R/2$, are used.
The main effect of this replacement is that there is no simple
relation between the units of density for this calculation and those
in the previous figures, hence the arbitrary units in
Fig.~\ref{fig:noneq}.
However, one may note that the form of the equilibrium boundary
(black) in Fig.~\ref{fig:noneq} is very similar to that in
Fig.~\ref{fig:phas1}, and one may thus estimate the size of shift
associated with pumping and decay.
Dephasing shifts the phase boundary to higher densities; this would
reduce the distance between the predicted phase boundary and the
critical density estimated from the blueshift.

\subsection{Condensate fraction}
\label{sec:condensate-fraction}

\begin{figure}[htpb]
  \begin{center}
    \includegraphics[width=0.95\linewidth,angle=0]{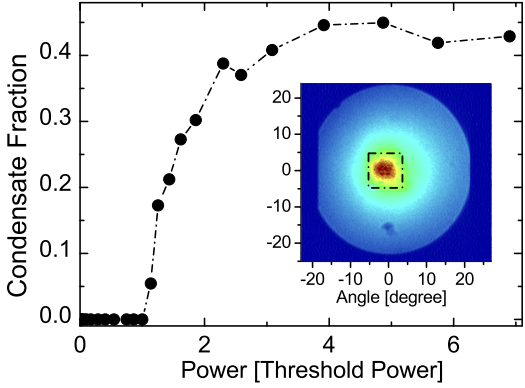}
  \end{center}
  \caption{(Color online) Experimental condensate fraction (the
    fraction of polaritons in the low momentum peak), at detuning
    $\delta=10.5$meV, $T_{\text{cryo}}=5.4$K, $T_{\text{eff}}=16.5K$.
    Inset: Occupancy vs angle, showing the peak at small angles (low
    momentum), the integrated weight inside which is taken as
    condensate density.}
  \label{fig:condfrac}
\end{figure}

As a further illustration of how dephasing depletes the condensate, we
compare experimental condensate fraction (Fig.~\ref{fig:condfrac})
with the theoretical quasi-condensate fraction (evaluated at the
mean-field level) with and without pumping and decay
(Fig.~\ref{fig:noneq}).
Although trapping negligably affects the phase boundary (the
properties of the phase boundary being dominated by interactions),
trapping and finite size are important in allowing a non-zero
condensate fraction\cite{petrov00,bayindir98}.
A quasi-condensate fraction can also be defined for an infinite
two-dimensional system; its definition follows from the behaviour of
the off-diagonal one-particle density matrix $\rho(r)$.
In three dimensions $\rho(r)$ tends to a constant at long distances,
which defines Off-Diagonal Long-Range order; in two dimensions it
instead decays as a power law.
However, the power law decay is due to phase fluctuations, and by
separating the effects of phase fluctuations from the quasi-condensate
depletion due to density fluctuations~\cite{kagan00} one may extract a
quasi-condensate density.
In a trapped two-dimensional system there can be a regime near the
phase transition in which there is quasi-condensation; however as
temperature is reduced or density increased, coherence will rapidly
extend over the entire system\cite{petrov00,keeling_review07}.
Thus, apart from a small region near the transition, coherence across
the entire system is expected even when the phase transition itself is
like the BKT transition.

The experimental condensate fraction is determined from the
occupancies illustrated in the inset of Fig.~\ref{fig:condfrac} as
follows:
The occupancy shows a clear peak at low momentum; the experimental
estimate of condensate fraction comes from the fraction of polaritons
found in this low momentum peak, i.e. the integrated density within
the FWHM of the occupancy peak, see
Ref.~\onlinecite{kasprzak06:thesis} for further details.
The theoretical mean-field estimate is the ratio of the coherent
density, $|\langle \psi \rangle|^2 + \sum_{\alpha} |\langle
S^-_\alpha\rangle|^2$ to the total (mean-field) density of
polaritons~\cite{szymanska06:long}.
There are of course complications in comparing the experimental
condensate fraction of a trapped 2D Bose gas\cite{bayindir98} to the
theoretical mean-field condensate fraction; however the comparison
between the various mean-field estimates clearly shows that pumping
and decay are responsible for a reduction of the
quasi-condensate fraction.

\section{Conclusions}
\label{sec:conclusions}

Concluding, we have  compared experimental and theoretical
phase boundaries for condensation of CdTe microcavity polaritons.
The experimental data lie near the crossover between a regime
  where the transition can adequately be described by
condensation of structureless bosons and a mean-field regime
where instead the long-range nature of the polariton-polariton
interaction determines the boundary.
We were limited to temperatures near this crossover
region: Because of the short polariton lifetime, the polariton
temperature is decoupled from that of the lattice. 
There are small differences between the various ways of estimating
experimental density; were one to neglect effects of disorder these
differences would have been an order of magnitude larger.
However, these estimates of experimental density lie close to the
theoretical phase boundary, and considering the effects of pump and
decay improves this agreement.

\begin{acknowledgments}
  JK, FMM and MHS would like to acknowledge the financial support of
  EPSRC, JMJK acknowledges financial support from Pembroke College,
  Cambridge.
\end{acknowledgments}


\appendix

\begin{figure}
  \includegraphics[width=0.9\columnwidth]{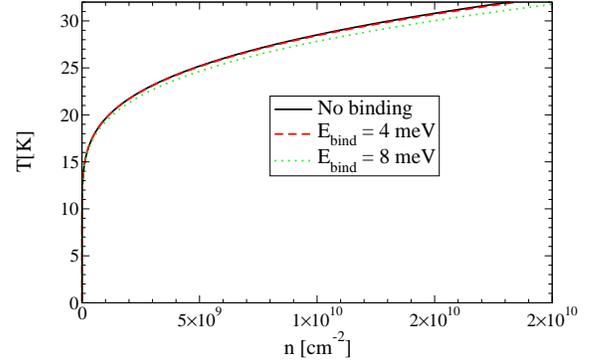}
  \caption{(Color online) Mean-field phase boundary calculated with
    varying strengths of biexciton binding energy, with a detuning
    $\delta = 5.3$meV (i.e. corresponding to top-left panel of
    Fig.~\ref{fig:phas1}). }
  \label{fig:biexcitons}
\end{figure}
%
\section{Interaction between spins and biexcitons}
\label{sec:inter-betw-spins}
This appendix discusses the effect on the phase diagram of including
an attractive interaction between opposite polarisations of two-level
systems.
The model used is a straightforward generalisation of
Eq.~\eqref{eq:hamil}, explicitly keeping track of the four possible
configurations of each disorder-localised site; unoccupied, singly
occupied by an exciton or occupied by a biexciton, thus each site
has an effective Hamiltonian:
\begin{equation}
  \label{eq:four-level}
  \hat{h}_{\alpha} =
  \left(
    \begin{array}{cccc}
      0 & \Lambda_{\alpha,L}^{} & \Lambda_{\alpha,R}^{} & 0 \\
      \Lambda_{\alpha,L}^{\ast} & \varepsilon_{\alpha} - \mu 
      & 0 & \Lambda_{\alpha,L}^{} \\
      \Lambda_{\alpha,R}^{\ast} & 0
      & \varepsilon_{\alpha} - \mu  & \Lambda_{\alpha,R}^{}  \\
      0 &  \Lambda_{\alpha,L}^{\ast} &  
      \Lambda_{\alpha,R}^{\ast} & 2(\varepsilon_{\alpha} - \mu) - \eb
    \end{array}
  \right)
\end{equation}
where $\Lambda_{(L,R),\alpha} = \sum_{\vect{k}} g_{\alpha,\vect{k}}
\psi_{(L,R)\vect{k}}/\sqrt{A}$ describes the coupling of a given
localised site to the left- or right-circularly polarised photon
field,
and $\eb$ is the biexciton binding energy.
This expression replaces the combination $(\varepsilon_{\alpha} - \mu)
S_{\alpha}^z + \sum_{\vect{k}}\left(g_{\alpha , \vect{k}}
  \psi_{\vect{k}}^{} S_{\alpha}^{+} + \text{h.c.}\right)/\sqrt{A}$
that appears in Eq.~\eqref{eq:hamil} for each site.
The mean-field theory, analogous to Sec.~\ref{sec:comp-exper-theor} is
found by calculating the mean-field estimate of density which may be written
as:
\begin{equation}
  \label{eq:four-level-density}
  \rho =\frac{1}{A} \sum_{\alpha}
  \frac{2 (z_{\alpha} + z_{\alpha}^2 \lambda)}{%
    1 + 2 z_{\alpha} + z_{\alpha}^2 \lambda},
\end{equation}
where we have defined:
\begin{displaymath}
  \lambda = e^{\beta \eb}, \qquad
  z_{\alpha} = e^{-\beta(\varepsilon_{\alpha} - \mu)},
\end{displaymath}
and the gap equation (found from the linear polarisability of the
four-level system) becomes:
\begin{equation}
  \label{eq:four-level-gap}
  \omega_0 - \mu
  =
  \sum_{\alpha}
  \frac{|g^{}_{\alpha,0}|^2/A}{1 + 2z_{\alpha} + z_{\alpha}^2 \lambda}
  \left[
    \frac{1-z_{\alpha}}{\varepsilon - \mu} 
    +
    \frac{z_{\alpha} (1-z_{\alpha} \lambda)}{\varepsilon - \mu - \eb}
  \right].
\end{equation}
It is straightforward to see that if $\eb=0$, these two forms
reproduce those in Eq.~\eqref{eq:two-level-density} and
Eq.~\eqref{eq:two-level-gap} respectively.
A more complete treatment of binding ought to take account of how the
binding strength varies according to the localised nature of the
single exciton states, however to place a bound on how large a shift
of critical temperature the biexciton binding might induce it is
sufficient to consider this model with a constant biexciton binding.

Figure~\ref{fig:biexcitons} shows the effect with binding energies up
to $8$meV, corresponding to the largest value seen for excitons bound
at vicinal surfaces in CdTe/CdMgTe wells~\cite{besombes00}. 
Such a value is almost certainly an overestimate for the samples used
in current experiment, as the results of Ref.~\onlinecite{besombes00}
are for narrower quantum wells than the current experiment ($3.2$nm vs
$5$nm), and correspond to biexcitons with the greatest in-plane
confinement.
Even with this overestimate of biexciton binding,
the effect of biexciton binding on the critical temperature can be
seen to be small.
Moreover, were biexcitons to play a significant role, one would expect
to see a crossover between the photoluminescence associated with
exciton-polaritons and that from biexciton-polaritons, as was
investigated in GaAs\cite{saba00}.
The fact that no such crossover is seen, and that the effects of
biexciton binding on the mean-field phase boundary are small justify
our neglect of biexciton physics in the rest of the paper.\newcommand\textdot{\.}



\end{document}